04

# COMPARACIÓN DE ALGORITMOS
PARA DETECCIÓN DE INTRUSOS EN ENTORNOS ESTACIONARIOS Y DE FLUJO DE DATOS

## A COMPARISON OF ALGORITHMS FOR INTRUDER DETECTION ON BATCH AND DATA STREAM ENVIRONMENTS.

MSc. Jorge Luis Rivero Pérez[1]
E-mail: jlrivero85@gmail.com
MSc. Bernardete Ribeiro[1]
E-mail: bribeiro@dei.uc.pt
MSc. Kadir Héctor Ortiz[2]
E-mail: khector@umet.edu.ec
[1] Universidad de Coimbra. Portugal.
[2] Universidad de Metropolitana. República del Ecuador.

## RESUMEN

La detección de intrusos en redes de computadoras a partir del enfoque de aprendizaje automático presenta algunas deficiencias dadas por la propia naturaleza de la aplicación. La principal viene dada por el modesto despliegue de sistemas de detección basados en algoritmos de aprendizaje bajo las restricciones impuestas por los entornos reales. En este artículo se describen y proponen tres variantes de pre procesamiento sobre el conjunto de datos KDD99, incluye selección de atributos. Luego la experimentación se realiza primeramente a partir de evaluar algoritmos representativos en entornos estacionarios sobre las variantes obtenidas a partir de pre procesar KDD99. Por último, dado que el tráfico de red es un flujo constante de datos, en el cual pueden existir variaciones de conceptos relacionadas con las tasas de falsos positivos, unido al hecho de que no se encuentran muchas investigaciones que aborden la detección de intrusos en entornos de flujos de datos nos conduce a realizar una comparación de varios algoritmos también representativos de flujos de datos. Como resultado se obtiene cuáles son los algoritmos que mejores resultados ofrecen en la detección de intrusos sobre las variantes de pre procesamiento propuestas, tanto para entornos estacionarios como de flujos de datos.

Palabras clave: Aprendizaje automático, detección de intrusos en redes, flujos de datos, KDD99.

## ABSTRACT

Intruders detection in computer networks has some deficiencies from machine learning approach, given by the nature of the application. The principal problem is the modest display of detection systems based on learning algorithms under the constraints imposed by real environments. This article focuses on the machine learning approach for network intrusion detection in batch and data stream environments. First, we propose and describe three variants of KDD99 dataset pre processing including attribute selection. Secondly, a thoroughly experimentation is performed from evaluating and comparing representative batch learning algorithms on the variants obtained from KDD99 pre processing. Finally, since network traffic is a constant data stream, which can present concept drifting with high rate of false positive, along with the fact that there are not many researches addressing intrusion detection on streaming environments, lead us to make a comparison of various representative data stream classification algorithms. This research allows determining the algorithms that better perform on the proposed variants of KDD99 for both batch and data stream environments.

Keywords: Data stream, KDD99, machine learning, network intrusion detection.

# INTRODUCCIÓN

En la actualidad la sociedad se va haciendo cada día más dependiente del uso de sistemas computarizados en diversas ramas como: las finanzas, la industria, la medicina y aspectos de la vida cotidiana entre otras. A su vez crecen las amenazas y los ataques lo que ha hecho que la Ciber Seguridad se convierta en un área de especial atención por parte de los especialistas, teniendo especial consideración en la capacidad de actuar pro-activamente con el objetivo de mitigar o prevenir los ataques. Dentro de esta área, la detección de intrusos es abordada desde enfoques estadísticos Marchete (2012) y de aprendizaje automático Garcia-Teodoro, Diaz-Verdejo, Maciá-Fernández & Vázquez (2009); Sangkatsanee, Wattanapongsa korn & Charnsripinyo (2011), Sommer & Paxson (2010); Tsai, Hsu, Lin & Lin (2009).

Los Sistemas de Detección de Intrusos en Redes (NIDS por sus siglas en inglés) son clasificados según sus métodos de detección. Los basados en firmas monitorizan la actividad comparándola con descripciones (firmas) de comportamientos maliciosos conocidos previamente; mientras que los basados en anomalías tiene la noción de actividad normal, clasificando como malicioso todo comportamiento desviado de ese perfil.

Varias son las investigaciones realizadas en la detección de intrusos en redes a partir de algoritmos de aprendizaje automático. Garcia-Teodoro, et. al (2009); Sangkatsanee, et. al (2011); Sommer & Paxson (2010); Tsai, et. al (2009). Pero a pesar de estas extensas investigaciones académicas, el despliegue de sistemas basados en aprendizaje automático para la detección de intrusos en ambientes operacionales se ha visto muy limitado (Sommer & Paxson, 2010). Esto ocurre debido a la propia naturaleza de la aplicación, la cual exhibe características particulares que hace que un despliegue efectivo sea más complicado que en otros contextos. Investigaciones previas han fundamentado lo anteriormente planteado identifica algunos aspectos que resultan claves, en los cuales los enfoques de aprendizaje automático no alcanzan su mejor rendimiento. Ejemplo de ello es la detección de patrones que no se ajustan a la distribución de los datos (outliers) ya que los algoritmos de aprendizaje automático en esencia ofrecen mejores resultados encontrando similitudes, o sea, en tareas de clasificación, que identifican actividades no se ajustadas a un patrón. Esto último es muy necesario en la detección de intrusos basada en anomalías. Por otra parte el costo relativo de una mala clasificación es extremadamente alto comparado con otras aplicaciones de aprendizaje automático. Un falso positivo requiere el consumo de mucho tiempo de los especialistas. Se examina el incidente reportado para eventualmente determinar que el mismo refleja una situación normal. Estudios argumentan que una tasa pequeña de falsos positivos puede inutilizar un NIDS (Modi, et. al., 2013). Además, los falsos negativos tienen el potencial para comprometer seriamente la integridad de la infraestructura informática y de comunicaciones.

En la comunidad de detección de intrusos se tiende a limitar la evaluación de los sistemas de detección de anomalías al cálculo de la desviación de las nuevas instancias respecto al perfil normal. Constituye un reto convertir sus resultados en reportes semánticos para los operadores de redes. Por lo general este último paso no es abordado por las investigaciones, es una carencia actual. Al sistema detectar situaciones anómalas, o sea que se desvían del perfil normal, los operadores de redes se hacen preguntas como: ¿Qué significa? Esa es la principal diferencia entre actividad anómala y ataque. Se puede afirmar que los sistemas de detección basados en anomalías reportan actividad que no ha sido vista nunca, la cual puede ser normal o no. Se hace necesaria una interpretación semántica de los resultados para el despliegue operacional de estos sistemas, ya que el objetivo es detectar ataques y por lo general la tasa de falsos positivos es muy alta. Por otra parte el tráfico de red resulta diverso, debido a que comúnmente características como el ancho de banda, la duración de las conexiones y la variedad de las aplicaciones muestran gran variabilidad. Esto hace que para los sistemas de detección de intrusos basados en anomalías sea difícil encontrar una noción estable de normalidad en el tráfico (Sommer & Paxson, 2010).

Otra cuestión que se considera atenuante para el despliegue de estos sistemas es que tradicionalmente la detección de intrusos a partir de aprendizaje automático se ha trabajado en entornos estacionarios, donde los datos permanecen disponibles en todo momento y son divididos, utilizando una porción para entrenar los algoritmos y otra para evaluarlos. Frameworks como WEKA (Bouckaert, et. al., 2013) son muy utilizados para estas tareas ya que implementan varios algoritmos para el aprendizaje así como métricas para evaluar y establecer comparaciones. La evaluación de algoritmos de aprendizaje en estos entornos para la detección de anomalías en redes, resulta útil como base para otras formas de descubrimiento del conocimiento como son los sistemas basados en reglas. Pero este enfoque se aleja del fenómeno real ya que el tráfico de red es un flujo constante de datos y para lograr actuar de manera proactiva se requiere de algoritmos capaces de aprender en tiempo real a partir de instancias de datos que van arribando en fracciones de tiempo muy pequeñas. Estos entornos de aprendizaje son los denominados flujos de datos, donde los datos no están idénticamente distribuidos por lo que existen variaciones de conceptos pudiendo constituir variantes de nuevos ataques (Gama & Gaber, 2007; Gama, Sebastião & Rodriguez, 2009; Shaker & Hüllermeier, 2012).

Para su mejor comprensión este artículo está dividido en secciones, en las que se describen algunos de los conjuntos de datos disponibles para la evaluación de propuestas de sistemas de detección de intrusos en redes de computadoras. Luego se presentan algunas variantes de preprocesamiento de los mismos. Por último se evalúan y comparan algoritmos de clasificación representativos de diferentes enfoques de aprendizaje automático tanto en entornos estacionarios como de flujos de datos, se utiliza para ello frameworks que implementan además de los algoritmos, metodologías de evaluación y métricas de comparación.

# DESARROLLO

En la presente investigación se realiza un estudio de diferentes variantes de preprocesamiento sobre el conjunto de datos KDD99. Luego se proponen tres variantes sobre las cuales se evalúan varios algoritmos representativos del aprendizaje automático tanto en entornos estacionarios como de flujos de datos. Para ello se han tenido en cuenta diferentes metodologías y métricas de evaluación bien establecidas para estas tareas. Las mismas permiten establecer una comparación confiable para determinar cuáles son los mejores resultados, en este caso en la detección de intrusos.

La investigación sigue una secuencia lógica y ordenada de etapas en la detección de intrusos. Se desarrolla una primera etapa en la que a partir del estudio y la experimentación de investigaciones previas en este campo de acción se logra proponer tres variantes de preprocesamiento. Luego son seleccionados y evaluados algoritmos representativos de diferentes enfoques dentro del aprendizaje automático en entornos estacionarios. De igual manera se seleccionan y evalúan algoritmos de entornos de flujos de datos.

Para llevar esta investigación a la práctica se utilizan los frameworks WEKA y MOA, para entornos de aprendizaje estacionarios y de flujos de datos respectivamente. Ambos implementan los algoritmos, metodologías y métricas antes mencionadas, se facilita así la reproducibilidad de los experimentos.

En la selección de los métodos se tuvieron en cuenta aspectos como: los datos que se necesitan obtener, la correspondencia con el diseño teórico y la estrategia investigativa seleccionada.

La relativa falta de conjuntos de datos de alta calidad para la detección de intrusiones es un problema en esta área. Debido a esto algunos investigadores han construido sus propios conjuntos de datos. Sin embargo, esta solución se enfrenta a la dificultad de etiquetar correctamente los mismos. Para ello se emplean varias herramientas como honey-pots[1] y honey-nets[2], combinadas con ataques para así lograr etiquetarlos de manera precisa, pero estos enfoques aun enfrentan varios retos. Además, el resto del tráfico no se puede asumir siempre como normal, ya que también puede estar contaminado con datos correspondientes a ataques. Otra cuestión es que los conjuntos de datos deben actualizarse constantemente con nuevas instancias al contener nuevo tráfico normal (correspondiente al uso de nuevas tecnologías, al despliegue de nuevas aplicaciones y a nuevos usuarios) y ataques (nuevas técnicas o vulnerabilidades) para entrenar interactivamente a los sistemas de detección de intrusos en la medida que evolucionan las nuevas tecnologías y los ataques. Conjuntos de datos públicos de alta calidad, robustos y diversos son fundamentales para estos problemas. Las investigaciones actuales referidas a la producción de los mismos facilitan a los investigadores tener un mejor progreso general en la detección de intrusos. Algunas fuentes sugieren que la detección de intrusos en algunos escenarios debe utilizar clasificación múltiple, es decir, utilizar más etiquetas (ataque, normal, sospechoso, desconocido, etc.) para caracterizar el tráfico, en lugar de usar clasificación binaria a partir de solo dos etiquetas (ataque, normal) (Sommer & Paxson, 2010).

Los primeros conjuntos de datos disponibles DARPA98 y DARPA99 han sido creados a partir de capturar el tráfico de red con TCPdump . Luego, basados en estos propusieron KDD99 (Rivero Pérez, 2014). Este se ha convertido en un estándar dentro de los conjuntos de datos de gran volumen para la evaluación de diferentes algoritmos de aprendizaje automático. Sobre el mismo se han desarrollado diversos estudios, los que han dado lugar a algunas variantes del mismo como son NSL-KDD y KDD99-10.

Otros conjuntos de datos recientes son: ISCX (Shiravi, Shiravi, Tavallaee & Ghorbani, (2012), MAWI (Fontugne, Borgnat, Abry & Fukuda, 2010). A pesar de ser más actuales, estos no son tan utilizados como KDD99 y sus variantes (Ibrahim, Basheer & Mahmod, 2013; Revathi & Malathi, 2013; Rivero Pérez, (2014). La experimentación desarrollada en este artículo, tanto en entornos estacionarios como de flujos de datos se realiza sobre KDD99. A continuación se realiza una breve descripción del mismo y se definen cuáles fueron los atributos seleccionados en la etapa de preprocesamiento de los datos.

---

1 Software o conjunto de computadores cuya intención es atraer a atacantes, simulando ser sistemas vulnerables o débiles a los ataques.

2 Tipo especial de *Honey-pots* de alta interacción que actúan sobre una red entera.

### Conjunto de Datos KDD99 y sus Variantes

KDD99 Consiste en registros de conexiones de red formados por 41 atributos. Los datos originales contienen 744 MB de 4 940 000. El conjunto de datos contiene 40 atributos por cada registro de conexión más otro atributo de etiquetado de la clase. Específicamente una conexión es una secuencia de paquetes TCP con un tiempo de inicio y fin bien definidos donde se enmarca el tráfico desde una dirección IP origen a una dirección IP destino a través de algún protocolo definido (Rivero Pérez, 2014).

En Song, Zhu, Scully & Price (2013), se explican los experimentos realizados para la conformación del mismo. Los 41 atributos que lo conforman se agrupan en las siguientes cuatro categorías (Rivero Pérez, 2014):

- Atributos básicos: se obtienen de los encabezados de los paquetes, sin inspeccionar el cuerpo del paquete. Son los 6 primeros atributos del conjunto de datos.

- Atributos de contenido: se obtienen a partir de un conocimiento del dominio aplicado al contenido del cuerpo de los paquetes TCP. Ejemplo: cantidad de intentos fallidos de inicio de sesión.

- Atributos de tráfico basados en tiempo: estos atributos fueron diseñados para capturar propiedades dentro de una ventana temporal de dos segundos. Por ejemplo el número de conexiones de una misma estación en un intervalo de dos segundos.

- Atributos de tráfico basado en estaciones: se utiliza una ventana histórica estimada a partir de un número de conexiones, en este caso 100. Estos atributos son diseñados para detectar ataques que sobrepasan los 2 segundos de duración.

KDD99 contiene alrededor de 5 millones de instancias, donde cada una representa una conexión TCP/IP que está compuesta por 41 atributos tanto cuantitativos como cualitativos. En muchas investigaciones se utiliza una pequeña porción que representa el 10 % del conjunto de datos original (variante conocida como KDD99-10), contiene 494021 instancias. Este subconjunto es utilizado para entrenamiento, mientras que para prueba se utiliza otro subconjunto que contiene 331029 instancias. Aproximadamente el 20% de ambos subconjuntos representan patrones normales de tráfico (no ataques). El conjunto de datos en su totalidad contiene 39 tipos de ataques agrupados en 4 categorías Rivero Pérez, (2014).

Algunas variantes han surgido a partir de KDD99. Ejemplo de ello es KDD99-10 contiene 22 tipos de ataques y es una versión más concisa que el conjunto original. Contiene más ejemplos de ataques que de conexiones normales. Debido a su naturaleza predominan los ataques del tipo DoS. La Tabla 1 muestra la cantidad de ejemplos de cada clase (Rivero Pérez, 2014).

Tabla1. Cantidad de instancias por clase en KDD99-10.

| Conjunto de datos | DoS | Probe | U2R | R2L | normal |
|---|---|---|---|---|---|
| KDD99-10 | 391458 | 4107 | 52 | 1126 | 97277 |

Entre las deficiencias de KDD99 (Kayacik, Zincir-Heywood & Heywood, 2005; McHugh, 2000; Tavallaee, Bagheri, Lu & Ghorbani, 2009), destaca el gran número de registros redundantes dado que aproximadamente el 78% y 75% de los registros en los conjuntos de datos de entrenamiento y de prueba del mismo se duplican. Esta gran cantidad de registros redundantes hace que los algoritmos de aprendizaje clasifiquen mejor las clases de las instancias más frecuentes, se dificulta el aprendizaje a partir de instancias poco frecuentes que son generalmente más perjudiciales para las redes, tales como ataques U2R. La existencia de estos registros repetidos en los conjuntos de prueba, hace que los resultados de la evaluación se inclinen por los métodos que tienen mejores tasas de detección, sobre los registros más frecuentes. En Tavallaee, et. al (2009), se proporciona una solución para resolver las cuestiones mencionadas, y se obtienen nuevos conjuntos de entrenamiento y prueba que constan de registros seleccionados de KDD99. La nueva variante creada, llamada NSL-KDD (Ibrahim et. al., 2013; Revathi & Malathi, 2013) no resulta redundante, cuenta con un total de 125973 instancias, lo que hace que sea asequible para realizar los experimentos en el conjunto de datos completo, sin necesidad de seleccionar al azar una pequeña porción para entornos estacionarios. Estas características, unidas al hecho de que los experimentos se realizan en una computadora personal de gama media, implica que los algoritmos no pueden analizar un volumen demasiado grande de datos, propicia el empleo de queNSL-KDD ,en este artículo para evaluar los algoritmos de aprendizaje en entornos estacionarios. Luego, debido a que quizás la redundancia de KDD99 puede ser una característica asociada al tráfico de redes. En las redes es común encontrar más tráfico normal que de ataques en un largo período de tiempo. Se hace considerar KDD99-10 como conjunto de datos para la evaluación de algoritmos en entornos de flujos de datos. Ambos conjuntos de datos cuentan con los mismos atributos y clases, por lo que se aplican las mismas variantes de preprocesamiento a ambos.

## Preprocesamiento de los Datos

Tanto el conjunto de datos KDD99 como sus variantes NSL-KDD y KDD99-10 tienen 23 clases, donde una clase es lo considerado como tráfico normal y las restantes 22 son consideradas ataques bajo cuatro categorías principales. Existen ataques de los cuales se tienen muy pocas instancias como el caso de spy (solo 2 instancias), perl (solo 3 instancias) y otras clases como la normal y smurf, las cuales cuentan con muchas instancias. En tal sentido, después de seleccionar los atributos previamente mencionados y para lidiar con esta situación se estudiaron las siguientes variantes de preprocesamiento de los datos (Rivero Pérez, 2014):

1. Modificar el conjunto de datos solo con la muestra de las cinco categorías como clases, las cuales serían: *normal, dos, probe, u2r, r2l*.

2. Modificar el conjunto de datos solo con la muestra de dos clases: *ataque* y *normal*. Sobre esta variante se aplica el algoritmo de máquinas de soporte vectorial *SMO* en la experimentación.

3. Mantener como etiquetas de clases los 23 tipos de ataques contenidos en el conjunto de datos.

La selección de atributos para esta investigación se realiza al tener en cuenta los atributos seleccionados como más relevantes en el estudio realizado sobre las diferentes técnicas de preprocesamiento, aplicadas sobre KDD99 (Rivero Pérez, 2014) y los obtenidos a partir de aplicar como algoritmo evaluador de atributos OneR AttributeEval con el método de búsqueda Ranker. Una vez fusionados los resultados fueron seleccionados los atributos: 1, 2, 5, 6, 9, 23, 24, 29, 32, 33, 34 y 36 (Rivero Pérez, 2014).

## Detección de Intrusos en Entornos Estacionarios

En esta sección se describen los resultados obtenidos al evaluar algoritmos representativos de los diferentes enfoques de aprendizaje automático en entornos estacionarios. Las evaluaciones se desarrollan con el framework WEKA (Bouckaert, et. al., 2013) que además de implementar la mayoría de los algoritmos de clasificación del estado del arte implementa diferentes metodologías de evaluación. En estos experimentos se utiliza validación cruzada con valor de 10 (crossvalidationfold 10) garantizando que cada instancia fuera utilizada al menos una vez para entrenar y otra para probar. Son evaluados algoritmos de diferentes enfoques de clasificación sobre el conjunto de datos NSL-KDD. A continuación se referencian trabajos relacionados con la aplicación de algoritmos de clasificación para la detección de ataques y se exponen los resultados obtenidos en esta investigación.

## Evaluación de los Algoritmos en Entornos Estacionarios

El primer algoritmo evaluado ha sido una Red Neuronal Perceptrón Multicapa (MLP). Esta es una de las redes neuronales más usadas para la clasificación. En Sabhnani & Serpen (2003), aplican MLP para detectar ataques en el juego de datos KDD99. Su arquitectura consiste en una red feed-forward de tres capas: una de entrada, una oculta y una capa de salida. En las capas oculta y de salida se utiliza la función Unipolar sigmoid con un valor slope de 1.0. El algoritmo de aprendizaje usado es el gradiente estocástico descendiente con una función de error de cuadrado medio. La capa de entrada está formada por 41 neuronas (una para cada atributo) y la capa de salida está formada por 5 neuronas (una para cada clase). En los resultados reportados en Sabhnani & Serpen (2003), muestran que el 88.7% de los ataques Probe son detectados. Se detectan un 97.2% de los ataques DoS, un 13.2% de los ataques U2R y 5.6% de los ataques R2L. En Bivens, Palagiri, Smith, Szymanski & Embrechts (2002), usan un MLP combinado con Mapas Auto-organizados para clasificar ataques.

En el presente experimento se mantiene la misma arquitectura de red neuronal que se referencia. Solamente varía la cantidad de neuronas de la capa de entrada. Se realizas una selección de atributos y no se trabaja con el total de 41. Además en dependencia de las variantes de preprocesamiento varia la cantidad de neuronas en la capa de salida pues la cantidad de clases es diferente para cada variante. En la primera han sido clasificadas correctamente 124109 instancias para un 98.49%, e incorrectamente 1864 instancias, lo que representa un 1.48%. En la variante 2 se obtiene un 98.58% de clasificación correcta sobre 124184 instancias y un 1.42% de clasificación incorrecta para 1789 instancias. Por último sobre la variante 3 los resultados arrojan un 98.49% de clasificación correcta sobre 124071 instancias y un 1.51% de clasificación incorrecta sobre 1902 instancias.

*Máquinas de Soporte Vectorial (SVM): en Li, et. al., (2012); Mukkamala, Sung & Ribeiro, 2005) aplican clasificadores basados en* kernel[3] a problemas de detección de anomalías en redes de computadoras. Evalúan el impacto del tipo de kernel y de los valores de los parámetros en la exactitud con que clasifica los ataques SVM. La exactitud varía con el tipo de kernel así como con los valores de los parámetros. Una vez ajustados apropiadamente estos valores se logran clasificar con gran exactitud los ataques. Los resultados obtenidos en KDD99 muestran

---

3 Funciones que permiten convertir lo que sería un problema de clasificación no lineal en el espacio dimensional original, a un sencillo problema de clasificación lineal en un espacio dimensional mayor.

que más del 99% de los ataques son detectados por este algoritmo se usan los 6 atributos más relevantes. Sin embargo los autores en su artículo no dan una descripción detallada sobre sus experimentos. En Sung & Mukkamala (2003), usan SVM como clasificador en su propuesta de arquitectura de 3 capas para la detección de intrusos. En la última capa usan SVM multi-clase como clasificador para 4 categorías: Probe, DoS, U2R y R2L, se obtienen como resultados: 99.16 %, 97.65 %, 76.32% y 46.53 %, respectivamente. Estos resultados son mejores que los ganadores de la competencia que evalua el conjunto de datos KDD99. Los resultados de los falsos positivos no están reportados ni analizados en su artículo.

En esta experimentación se evalúa el algoritmo de SVM implementado en WEKA, conocido como SMO. Solo ha sido evaluada la variante 2 pues ese algoritmo es para tareas de clasificación binaria. Como resultado se obtuvo que 125242 instancias fueron clasificadas correctamente, lo que representa un 99.42% de clasificación correcta, mientras que 731 instancias resultan mal clasificadas lo que representa un 0.58% del total.

*Árboles de decisión: se estudia la propuesta de Sindhu, Geetha & Kannan (2012), y se evalúa el algoritmo de árboles de decisión implementado en* WEKA conocido como *J48*. Sobre la primera variante mencionada la clasificación se realiza sobre las clases *Probe*, *DoS*, *Normal*, *U2R* y *R2L*. De un total de 124738 instancias, se logra una clasificación de 125847 instancias correctas para un 99.02%, y 1235 instancias incorrectas para un 0.98%. En la variante 2 se obtiene un 97.43% de clasificación correcta sobre 122735 instancias y un 2.57% de clasificación incorrecta para 3238 instancias. Luego, sobre la variante 3 se logra clasificar correctamente 120745 instancias, lo que representa un 95.85% y 5228 instancias fueron clasificadas incorrectamente para un 4.15%.

Sobre esas mismas variantes se evalúa también el algoritmo *NaiveBayes*, donde en la primera variante de un total de 125973 instancias, se logra una clasificación correcta de 123642 instancias para un 98.15%, y 2331 instancias incorrectas para un 1.85%. En la variante 2 se obtiene un 98.5% de clasificación correcta sobre 124083 instancias y un 1.5% de clasificación incorrecta para 1890 instancias. Luego, la variante 3 logra una clasificación correcta de 121224 instancias de un total de 125973, para un 96.23%, y clasifica incorrectamente 4749 instancias lo que representa un 3.77%.

Luego se estudian variantes propuestas de aprendizaje basado en instancias para la detección de intrusos (Garcia-Teodoro, et. al., 2009) y se evalúa el algoritmo de clasificación basado en instancias *k* vecinos más cercanos (*k-NN*) implementado como *IBK* en WEKA sobre las diferentes variantes de preprocesamiento propuestas. Para ello se realiza la evaluación de diferentes valores de *k* (3, 5, 7, 9, 11 y 13). Sobre la variante 1, con *k*=3 han sido clasificadas correctamente 123984 instancias para un 98.42%, e incorrectamente 1989, lo que representa un 1.58%. En la variante 2 para ese mismo valor de *k* se obtiene un 98.02% de clasificación correcta sobre 123479 instancias y un 1.98% de clasificación incorrecta para 2494 instancias. Por último sobre la variante 3 los resultados arrojan un 97.94% de clasificación correcta sobre 123378 instancias y un 2.06% de clasificación incorrecta sobre 2595 instancias.

Con *k*=5 fueron clasificadas correctamente 123938 instancias para un 98.38%, e incorrectamente 2035 de las mismas, lo que representa un 1.62%. En la variante 2 se obtuvo un 97.98% de clasificación correcta sobre 123428 instancias y un 2.02% de clasificación incorrecta para 2545 instancias. Por último sobre la variante 3 los resultados mostraron un 97.89% de clasificación correcta sobre 123315 instancias y un 2.11% de clasificación incorrecta sobre 2658 instancias.

Con *k=7* fueron clasificadas correctamente 123783 instancias para un 98.26 %, e incorrectamente 2190, lo que representa un 1.74%. En la variante 2 para ese mismo valor de *k* se obtiene un 97.76% de clasificación correcta sobre 123151 instancias y un 2.24% de clasificación incorrecta para 2822 instancias. Por último sobre la variante 3 los resultados muestran un 97.86% de clasificación correcta sobre 123277 instancias y un 2.14% de clasificación incorrecta sobre 2696 instancias.

En la Tabla 2 se resumen los resultados de clasificación obtenidos por los algoritmos evaluados sobre las diferentes variantes de preprocesamiento. A partir de la misma puede hacerse un análisis que permite fusionar los resultados de los algoritmos con las diferentes variantes de preprocesamiento. Para la primera variante se obtiene mejor clasificación por parte del algoritmo de árboles de decisión J48, mientras que para la variante 2 resulta el algoritmo SMO de máquinas de soporte vectorial. En la variante 3 la mejor clasificación se obtiene con la Red Neuronal Perceptrón Multicapa (MLP).

Tabla 2. Resultados de clasificación sobre las diferentes variantes de preprocesamiento.

|  | 1 | 2 | 3 |
|---|---|---|---|
| MLP | 98.52% | 98.58% | 98.49% |
| SMO | - | 99.23% | - |
| J48 | 99.02% | 97.43% | 95.85% |
| Naive Bayes | 98.14% | 98.5% | 96.23% |
| k-NN (k=3) | 98.42% | 98.02% | 97,94% |
| k-NN (k=5) | 98.38% | 97.98% | 97.89% |
| k-NN (k=7) | 98.26% | 97.76% | 97.86% |

Detección de Intrusos en Entornos de Flujos de Datos

La experimentación en entornos estacionarios provee los criterios para el preprocesamiento de los datos, así como las bases para nuevas formas de extracción de conocimiento como es la generación de reglas a partir de los resultados obtenidos por algoritmos de árboles de decisión para su posterior implementación en sistemas basados en conocimiento. Sin duda resulta de gran utilidad pero dado el volumen de datos a analizar en tareas de detección de intrusos, en el que el tráfico de red es constante no se puede almacenar debido al gran volumen de información que implicaría, hace que metodologías de evaluación que dividen los datos para entrenamiento y pruebas como validación cruzada empleada anteriormente inmanejable pues los algoritmos solo pueden acceder a los datos una vez. Es decir, en la medida que arriban las instancias de tráfico. Esas características son propias de entornos de flujos de datos (Gama & Gaber, 2007; Gama, et. al., 2009; Shaker & Hüllermeier, 2012), además de estar caracterizados por datos arribando constantemente incluso de diversas fuentes, con tendencia a ser infinito, ocurren variaciones de conceptos que son un cambio en la distribución caracterizada por la generación de los datos. De ahí que los algoritmos para estos entornos además de tener la capacidad de procesar los datos en tiempo real deben detectar la ocurrencia de variaciones de concepto e implementar mecanismos para su tratamiento.

Con el objetivo de acercar esta investigación a la realidad de la aplicación en cuestión, a partir del despliegue de la misma en entornos de reales, al ser una de las deficiencias de la implementación del enfoque de aprendizaje automático para la detección de intrusos, comparamos en esta sección algoritmos de clasificación de flujos de datos para la detección de intrusos.

Para los entornos estacionarios existen varias metodologías que permiten la evaluación y comparación de algoritmos de aprendizaje incluso en múltiples conjuntos de datos o variantes del mismo (Demsar, 2006). En el caso de los algoritmos de clasificación en entornos de flujos de datos han sido propuestas algunas metodologías y métricas para evaluar el rendimiento de los clasificadores (Bifet, Read, Žliobait, Pfahringer & Holmes, 2013; Gama, et. al., 2009; Gama, Sebastião & Rodrigues, 2013; Shaker & Hüllermeier, 2012). Existen dos metodologías de evaluación para estos entornos conocidas como: *holdout* y *prequential*, las mismas son combinadas con mecanismos de olvido como: ventanas deslizantes y factores de desvanecimiento, que son requeridos para una rápida y eficiente detección de variaciones de conceptos. Estudios comparativos sobre las metodologías antes referidas han defendido el uso de prequential con factores de desvanecimiento como mecanismo de olvido (Gama, et. al., 2009, 2013), se demuestran las ventajas de la misma para calcular el rendimiento de los algoritmos. Para ello se calculan como métricas: la tasa de error y la exactitud. A continuación se describen los resultados obtenidos a partir de evaluar varios algoritmos de clasificación de flujos de datos para la detección de intrusos en tiempo real.

Evaluación de los Algoritmos en Entornos de Flujos de Datos

Para la evaluación de los algoritmos en entornos de flujos de datos se utiliza el framework MOA (Massive Online Analysis) (Bifet, Holmes, Kirkby & Pfahringer, 2010). Se mantienen todos los valores que tienen los parámetros por defecto. Se utiliza la metodología prequential en la cual cada instancia es evaluada antes de que el algoritmo entrene con ella, se asegura que todas son utilizadas una vez para entrenar y otra para clasificar. Como mecanismo de olvido se utiliza un factor de desvanecimiento con un valor de 0.95 se garantiza así un tratamiento adecuado ante las variaciones de concepto en el tráfico.

Todas las experimentaciones se realizan sobre la variante 2 de preprocesamiento pero en el conjunto de datos KDD99-10. Todas las experimentaciones se realizan sobre la variante 2 de preprocesamiento pero en el conjunto de datos KDD99-10. Los algoritmos evaluados fueron (Hoeffding Tree Rutkowski, Pietruczuk, Duda & Jaworski, 2013): HoeffdingTree, IBL Stream, Naive Bayes, Ozaa Boost. A continuación se muestran gráficamente los resultados en cuanto a exactitud de clasificación obtenidos por cada uno. Primero se muestran los resultados individuales y luego se muestra una gráfica que contiene todos los resultados:

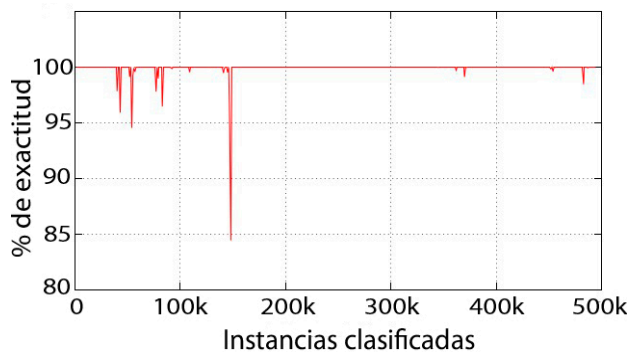

Figura 1. Resultados de la evaluación del algoritmo HoeffdingTree.

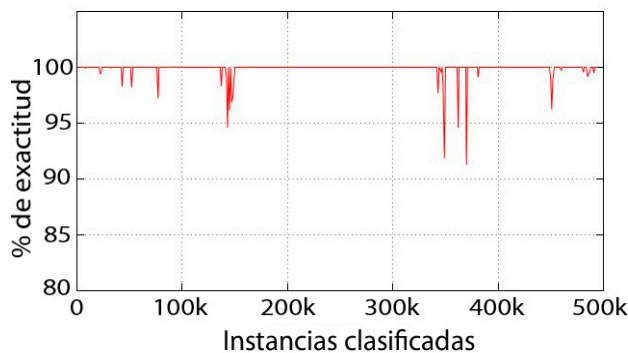

Figura 2. Resultados de la evaluación del algoritmo IBLStream.

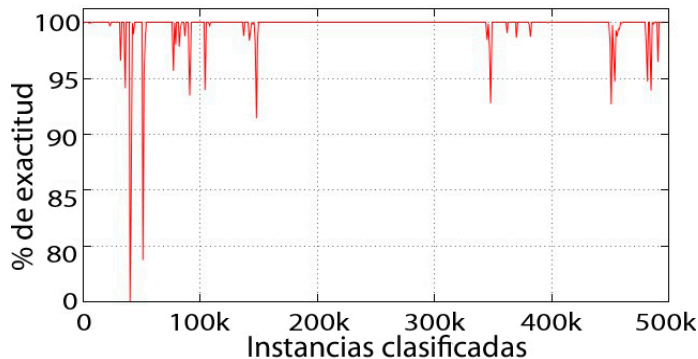

Figura 3. Resultados de la evaluación del algoritmo Naive Bayes.

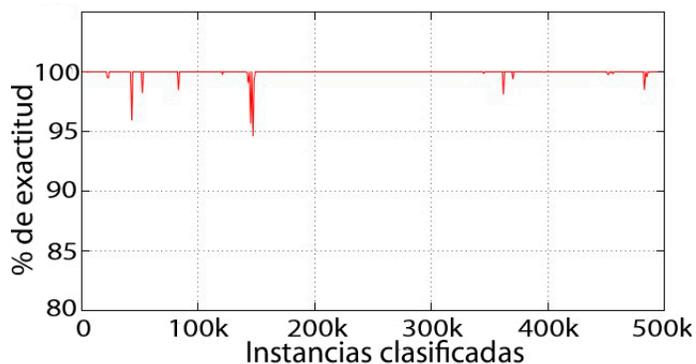

Figura 4. Resultados de la evaluación del algoritmo Ozaa Boost.

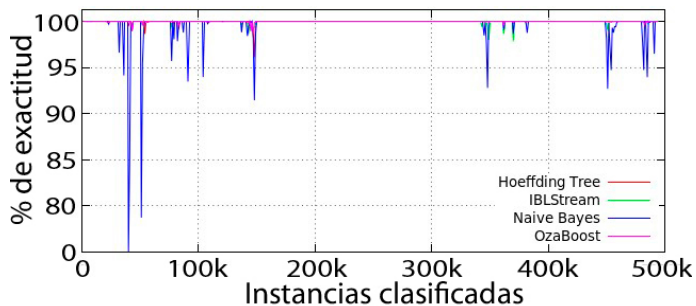

Figura 5. Comparación de todos los resultados.

A partir del análisis de las mismas se puede apreciar que existen variaciones de conceptos en las instancias 50788, 58628, 73274 y 150925 aproximadamente. Todos los clasificadores evaluados las detectaron, pero no se recuperaron en igual medida ante las mismas. La meta clasificadora OzaBoost ofrece los mejores resultados con la recuperación de las variaciones de concepto como en el promedio de la exactitud de la clasificación. De esa manera se logra corroborar en otras investigaciones que muestran la superioridad de estos algoritmos (Zhang, Zhu, Shi, Guo & Wu, 2011); también en la detección de intrusos los resultados son relevantes.

Tabla 3. Media de clasificación correcta de los algoritmos evaluados en entornos de flujos de datos.

| Algoritmos | Media de clasificacióncorrecta |
|---|---|
| Hoeffding Tree | 99.64 % |
| IBLStream | 99.81 % |
| Naive Bayes | 99.18 % |
| OzaaBoost | 99.87 % |

# CONCLUSIONES

La detección de intrusos bajo el enfoque de aprendizaje automático tiene varias deficiencias dada la naturaleza de la propia aplicación. A pesar de ello los investigadores continúan trabajando en lograr soluciones que permitan cubrir las mismas. Algo que resulta fundamental para el despliegue de estas soluciones es lograr que cada red, como sistema autónomo, haga la construcción de su propio conjunto de datos, lo cual debe actualizarse periódicamente debido a la diversidad de aplicaciones y al emergente crecimiento de las mismas, lo que puede provocar que el tráfico normal se clasifique cado como algún tipo de ataque. Así mismo surgen nuevos ataques y/o variantes de los ya conocidos. Existen varios conjuntos de datos disponibles para la evaluación de los algoritmos pero ninguno logra caracterizar de manera general el tráfico de las redes. KDD99 y sus variantes son los más empleados en la investigación científica.

Existen varias soluciones comerciales para la detección pero la mayoría son basadas en firmas, pero solo detectan ataques que ya tengan registrados. Lo adecuado es aplicar un enfoque de detección de anomalías a partir de determinar cuál es el algoritmo que mejores resultados ofrece sobre el conjunto de datos creado dentro del sistema autónomo. Para ello la evaluación de algoritmos en entornos estacionarios, a pesar de carecer de aplicabilidad cuando se refiere al fenómeno de tráfico en la red, requiere de menor infraestructura de cómputo, crea las bases para un criterio en cuanto a que algoritmos pueden resultar mejores, cuáles pueden ser las mejores variantes de procesamiento y cuáles serían las reglas de clasificación. Sirve de base para el aprendizaje automático a partir de flujos de datos de tráfico de red a pesar de que las metodologías de evaluación son diferentes. Luego, a partir del *framework* MOA puede construirse una solución personalizada para la red en cuestión, con la posibilidad de evaluar más de un algoritmo, incluso pudiendo desarrollar nuevas variantes de los mismos.

# REFERENCIAS BIBLIOGRÁFICAS